\documentclass{article}

\usepackage[pdftex]{graphicx}

\usepackage[separate-uncertainty=true,table-align-uncertainty=true]{siunitx}

\usepackage{amsmath}

\usepackage{url}
\usepackage{hyperref}
\hypersetup{
  colorlinks   = true, 
  urlcolor     = blue, 
  linkcolor    = blue, 
  citecolor   = blue 
}

\begin{document}

\title{Measuring Blackbody Noise in Silica Optical Fibres for Quantum and Classical Communication}

\author{Michael~Hencz \footnotemark[1],
        Mark~Baker\footnotemark[2],
        and~Erik~W.~Streed\footnotemark[1]\footnotemark[3]
\thanks{M. Hencz and E. W. Streed were with the Centre for Quantum Dynamics, Griffith University, Brisbane 4211 Australia.}
\thanks{M. Baker was with the Quantum Technologies, Sensors and Effectors Division, Defence Science and Technology Group.}
\thanks{E.W. Streed was with the Institute for Glycomics, Griffith University, Gold Coast, 4215 Australia.}}

%
%

\markboth{Journal of \LaTeX\ Class Files,~Vol.~14, No.~8, August~2015}%
{Shell \MakeLowercase{\textit{et al.}}: Bare Demo of IEEEtran.cls for IEEE Journals}
%



\maketitle

\begin{abstract}
    Deployment of practical quantum networks, which operate at or near single photon levels, requires carefully quantifying noise processes. We investigate noise due to blackbody radiation emitted into the guided mode of silica single mode optical fibres near room temperature, which to date is under-explored in the literature. We utilise a single photon avalanche detector and lock in detection to measure $\approx$0.1 photons/s/THz ($\approx$-170dBm/THz) at 40°C near the optically thick limit of 20km in silica fibre. We also measure a coarse spectrum to validate the blackbody behaviour, and observe a prominent anomaly around the 1430nm CWDM channel, likely due to -OH impurities. Though the magnitude of this noise is small, it is additive noise which imposes a fundamental limit in raw fidelity in quantum communication, and a fundamental noise floor in classical communication over optical fibres.
\end{abstract}

\section{Introduction}
\begin{figure}
    \includegraphics[width=\linewidth]{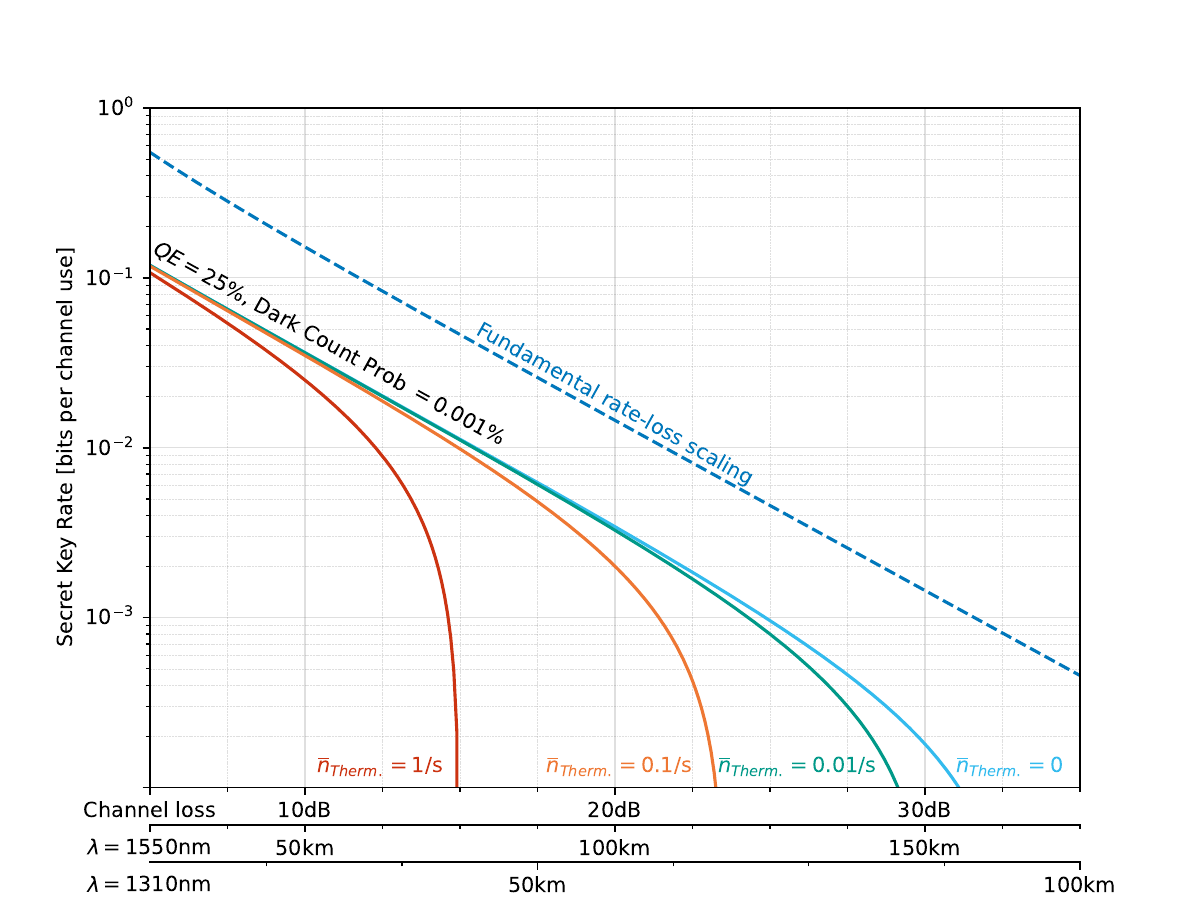}
    \caption{Secret key rate calculated following the fundamental rate-loss limit with thermal noise presented in \cite{RN636}. This has been modified to show the expected rate for a channel using a real SPAD detector (Quantum efficiency=25\%, Dark Count probability = 0.001\%). The thermal noise has been modeled as constant rates $\bar n _{\text{therm.}}$, reduced by a detector duty cycle of 1/1000 (which would be found in a 1MHz generation rate with 1ns trigger window). Based on our measurements at 40\si{\celsius}, $\bar n _{\text{therm.}}=0.1$\si{\per\second} is expected for a 1\si{\tera\hertz} bandwidth.}
    \label{Fig:RateLoss}
\end{figure}

%
%
%
%
Understanding fundamental limits in quantum communication is critical for the design and deployment of large-scale quantum networking, for both quantum communication and computation applications. The long-distance rate-loss limit has been investigated theoretically \cite{RN636} and is shown in Figure \ref{Fig:RateLoss}, though some QKD protocols allow rates in excess of this limit \cite{RN673}. This rate-loss limit is generally understood in the context of attenuation, resulting from the $\approx0.14$ to $\approx0.18$ dB/km loss introduced from the use of silica optical fibres as the transmission medium \cite{RN780}, and is dominated by Rayleigh scattering in the optical telecommunications band (1300-1600\si{\nano\meter}). In the single photon level signals present in quantum communication, this attenuation corresponds to the probability of losing the signal photon across the communication channel.

Another important aspect to the long-distance limit is the presence of noise in a quantum channel \cite{RN636}. Single photons of background noise can disrupt the quantum channel if present in the timing, spectral, and polarisation detection window. Background noise due to cross-coupling from classical signals on adjacent channels is well understood \cite{RN119, RN677}. However, blackbody noise emitted by silica fibre is not well investigated experimentally, despite the widespread use of optical fibres for thermal sensing in higher temperature regimes \cite{RN660, RN668}. Figure 1 indicates our calculation of the impact of thermal noise based on the methods presented in \cite{RN636}.

Blackbody radiation emitted into the guided mode of single mode optical fibre has the potential to be highly transmissive, and due to the random nature of thermal emission, difficult to mitigate in silica fibre. Additive background noise imposes a limit on the maximum fidelity achievable for a given quantum channel over an optical fibre of a given distance, and impose a fundamental noise floor for classical communication.

Molecular vibrations from -OH impurities presents a potential noise source in addition to the blackbody emission. While the reduction of residual -OH has been a major technical accomplishment and low -OH fibre is standardised, the -OH absorption peak at 1383\si{\nano\meter} is still specified as contributing 0.4dB/km in low -OH fibre\cite{RN71}. Though this is within the telecom band, these are overtones of fundamental resonances \cite{RN666} in the mid IR region ($\sim$ 2800\si{\nano\meter}), where silica is highly opaque ($\sim$$10^5$\si{\decibel\per\kilo\meter} range). Absorption of blackbody radiation in the mid IR region thus has a mechanism for molecularly driven re-emission spectrally proximate to the telecom bands. Measurements of this emission have been conducted at temperatures ranging from 600 - 1100°C \cite{RN667}, but measurements near room temperatures are more relevant for practical networking, and extrapolation of high temperature results to ambient temperatures is challenging.

In this work, we will discuss our techniques for measuring the blackbody radiation emitted into the guided mode of standard single mode telecom fibre at temperatures ranging from 40°C to 180°C, with higher temperatures used to validate the trend in the data for extrapolation to temperatures lower than measured. We show that there is a small but unavoidable amount of thermal background noise present in the commonly used channels around the minimum in the attenuation spectrum of silica, which we suspect is enhanced by the hydroxide impurities. We also show that the blackbody radiation is fibre length dependant.

\section{Method}
\subsection{Theory}
Planck's law describes the power spectral density of thermal radiation from an ideal (perfectly emissive) surface for a unit surface area at all emission angles \cite{RN680}. For transparent media, there is a requirement of optical thickness \cite{RN679} for blackbody emission, $\tau \gg 1$, where $I/I_0 = e^{-\tau}$. In optical fibres, the $\tau=1$ threshold corresponds to approximately $-4$\si{\decibel} or $\approx20$\si{\kilo\meter} of fibre for 1550\si{\nano\meter} (with other wavelengths being optically thick at shorter distances). This suggests that the thermal emission for optical fibre has the potential to exhibit non-trivial behaviour for typical distances in fibre-based long distance quantum communication. The emissivity of silica can be calculated using Kirchoff's law of thermal radiation, which states that the emissivity and absorptivity of a material must be equal \cite{RN680} as they are symmetrical processes. This means that distances longer than 20\si{\kilo\meter} will emit more blackbody light in the telecom band for a given temperature, trending towards ideal blackbody behaviour as the transmission trends to zero at long distances.

The losses in silica fibre are well investigated \cite{RN55}. The three absorptive contributions are: IR vibrational losses dominant at longer wavelengths; the Urbach direct bandgap absorption dominant at shorter wavelengths; and impurity absorption at specific wavelengths including the water peak (which absorbs around 1390-1400\si{\nano\meter}). Rayleigh scattering dominates the attenuation spectrum for telecom wavelengths but will not contribute to the emissivity, as it is not an absorptive loss. To estimate an expected signal level in our experiment, we use a model of a silica fibre blackbody system using Planck's law and the theoretical absorptivity of impurity free silica fibre. The details of these calculations are discussed Appendix \ref{App:Theory}. Importantly, these calculations suggest we will be crossing a signal level on the order of 1 photon per second at approximately 45°C across the bandwidth of our detector when accounting for the quantum efficiency, and as such we have designed our experimental method around this. These calculations will also allow the determination of the noise for longer lengths of fibre.

\subsection{Experiment}

We tested four spools of commercial optical fibre: G652B 20\si{\kilo\meter}; G652B 3\si{\kilo\meter}; G652D 20\si{\kilo\meter}; G652D 2\si{\kilo\meter}. The G652B (High -OH) standard is the older standard which doesn't have a specification of the maximum permissible water peak (though there is an implicit maximum due to the impact on specified wavelengths). The G652D (Low -OH) standard has a maximum permissible water peak specified at 0.4\si{\decibel\per\kilo\meter} at 1383\si{\nano\meter}. The two different standards were selected to determine if there was any appreciable difference in the blackbody emission characteristics between low and high -OH fibre. The two different lengths were selected to give insight into the optical thickness threshold, as we expect significantly less light for the shorter distance, however the theoretical model may not be accurate due to the length being less than the optical thickness for telecom wavelengths.

\begin{figure}
    \includegraphics[width=\linewidth]{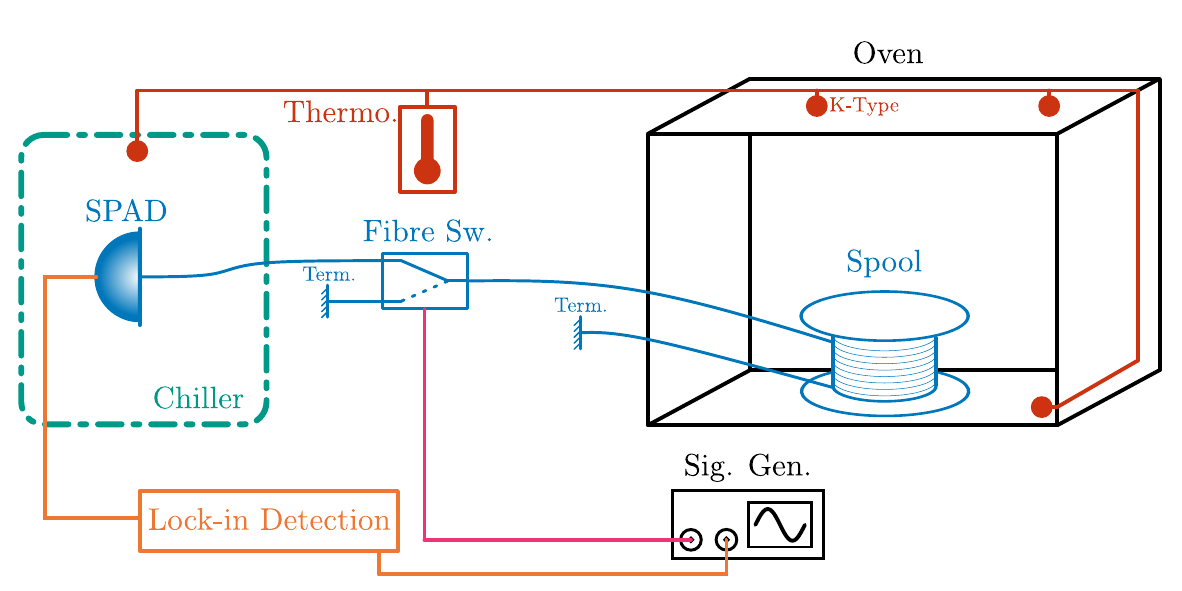}
    \caption{Configuration used for measurement. The spool under test was placed in an oven, with temperatures recorded using K-type thermocouples. The SPAD used for lock-in detection was placed in a temperature controlled chilled box to reduce dark count rates.}
    \label{Fig:ExperimentConfig}
\end{figure}

Heating of the spool under test is achieved using a temperature controlled lab oven (Yamoto DH411). The temperature was measured using three K-type thermocouples positioned as in Figure \ref{Fig:ExperimentConfig}, connected to a thermometer with a USB interface for data logging. To account for thermalisation time, we only use data obtained after approximately 30 minutes at a given temperature. Our measured temperature variation during each measurement interval was on the order of 1°C. The splices for the pigtail ends of the fibre spool under test were run out of the oven via a rear cable conduit. To avoid cladding mode propagation, we wound 15 turns of fibre around a 1 inch mandrel after the splice in the pigtail ends. Our ambient temperature was $\approx$21°C, and we chose an upper temperature of 180°C to avoid thermal decomposition of the ABS spool and acrylate coating (estimated to occur near 200°C \cite{RN684}).

We used an ID Quantique Qube NIR SPAD operating in free running mode with 20\% quantum efficiency to coincide with the efficiency curve given by the manufacturer. We set a 40µs deadtime to minimise the amount of after-pulsing, and we measured an after-pulsing probability $A\approx 6 \times 10^{-6}$, using the power law curve fit discussed in \cite{RN783} fit to time tag data with the detector free running. This model of SPAD is cooled by an internal TEC to -30\si{\celsius}, but was observed to be sensitive to air temperature, likely as a result of temperature gradients across the active region. This manifests as an increase in the dark count rate dependant on the air temperature. To account for this, we placed the SPAD in a chilled, temperature stabilised box. This stabilised the temperature to $16.5\pm0.5$\si{\celsius} with a corresponding dark count rate of $\approx$2200\si{\per\second}, with a temperature sensitivity of $\approx62$\si{\second\per\kelvin} at this temperature.

To overcome the noise floor imposed by the dark count rate, we implemented a lock-in detection scheme to allow us to statistically measure the presence of signals with magnitudes much smaller than our raw detector noise floor. For small signals, this scheme should also average out detector after-pulses, as they will be randomly distributed in time. This scheme is similar to the work in \cite{RN682}. Using lock-in detection, we can estimate the average number of true signal counts in excess of our detector noise floor by modulating the measured signal. Our modulation is achieved by using an optical fibre MEMS switch running at 27Hz to avoid mains related noise processes. This allows us to measure true signal count rates down to $\approx$1\si{\per\second} after integrating for approximately 8 hours, giving an approximately 33\si{\decibel} improvement. Our integration time varied between 4 hours for temperatures above 100\si{\celsius}, and 14 hours for measurements around 40\si{\celsius}. For longer integration times, we found that the increased signal sensitivity was not worth the time investment required. Simulation of the measurement technique found that an uncertainty of $\pm2 \sigma$ in the curve fit was a good approximation for the simulated measurement error.

\section{Discussion}
\begin{figure}
    \includegraphics[width=\linewidth]{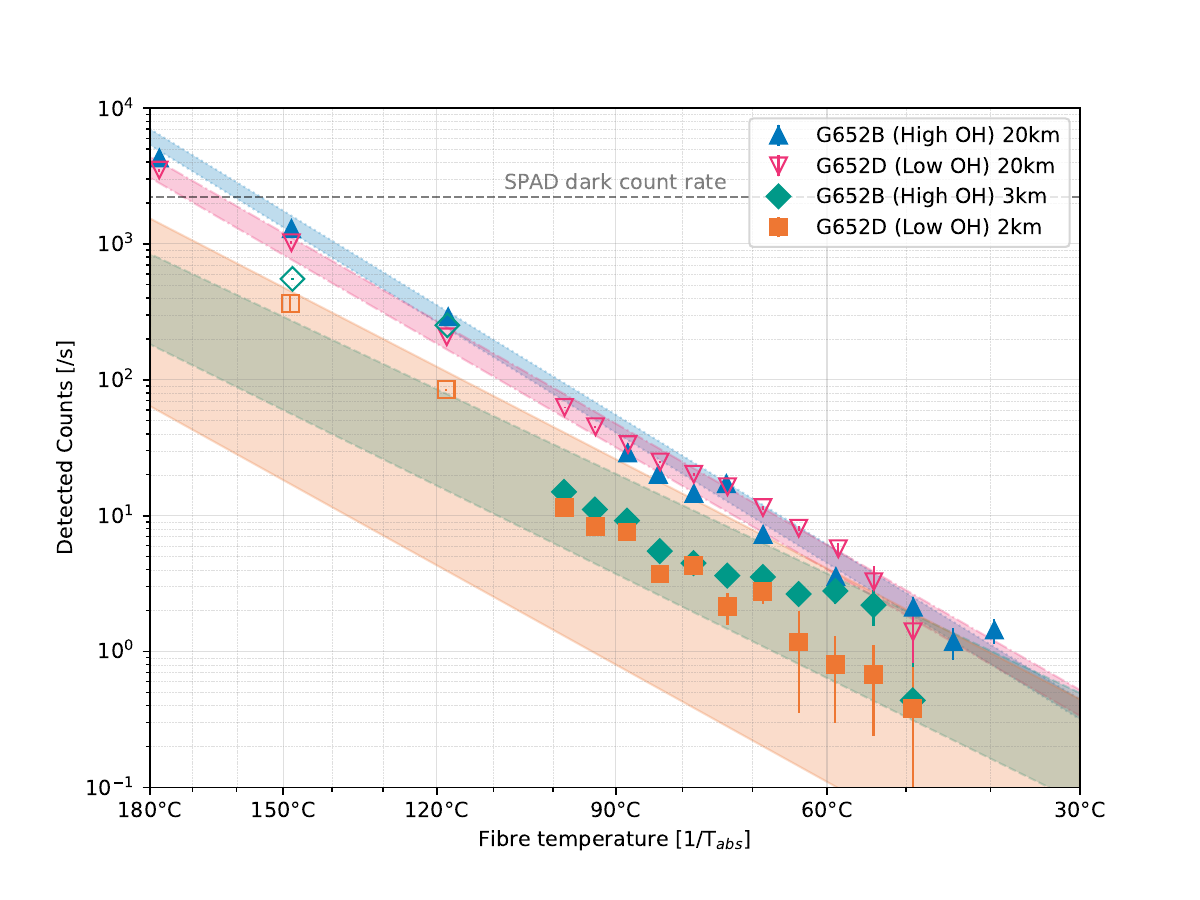}
    \caption{Blackbody emission measured using lock-in detection of the four telecom fibres under test. Solid points are obtained from intact fibre, outlined points indicate possible fibre breakage before or during measurement. Curve fit results are summarised in Table \ref{Tab:CFResults}, with fit errors indicated by the shaded area.}\label{Fig:BBResult}
\end{figure}
\begin{figure}
    \includegraphics[width=\linewidth]{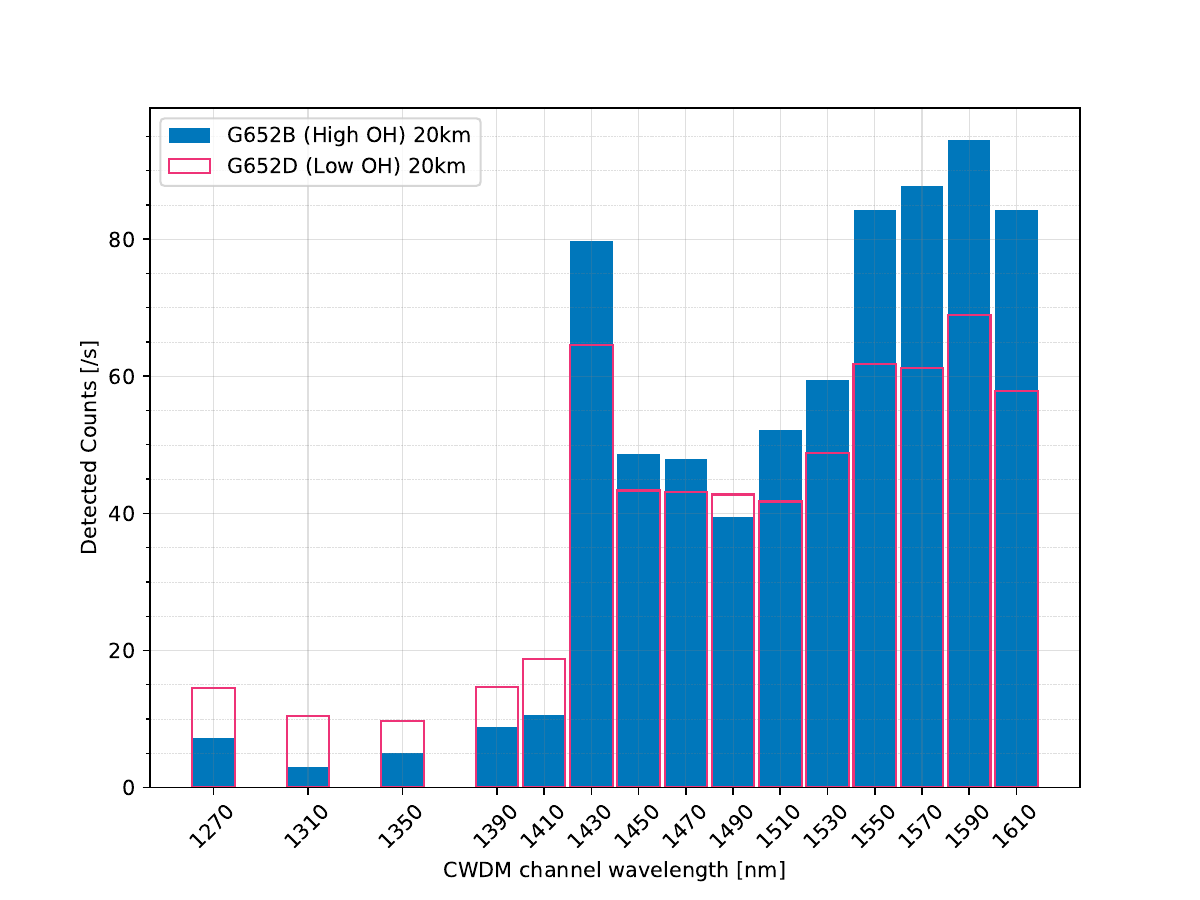}
    \caption{Measured spectrum in telecom channels of light emitted at 150°C measured using a CWDM multiplexer. G652D results are obtained from a broken fibre.}.
    \label{Fig:SpecResult}
\end{figure}

Overall, the behaviour of the emission with respect to temperature is as expected: Higher temperatures result in an exponential increase in the blackbody emission, which follows the experimental data down to 40\si{\celsius}, with a rate of approximately 1 \si{\per\second} (equivalent to a power of -158dBm) across the entire detection range shown in \autoref{Fig:BBResult}, with the result in the 1550nm channel suggesting a noise level of 0.1\si{\per\second\per\tera\hertz} at 40\si{\celsius} for 20\si{\kilo\meter} shown in \autoref{Fig:SpecResult}. For 20\si{\kilo\meter} lengths, we obtain reasonable agreement with our theoretical model, with curve fitting results shown in \autoref{Tab:CFResults}, and lower lengths emit significantly less light as expected.

\begin{table}
\centering
\caption{ Curve fitting for results presented in Figure \ref{Fig:BBResult}, and theoretical calculations from \autoref{App:Theory}. Fit equation is $A \exp{\left( \frac{hc}{k_B}\frac{1}{\lambda_c T}\right)}$, with $T$ in units of K, and plots of the theoretical fits can be found in \autoref{Fig:TheoryCurves}.}\label{Tab:CFResults}
      \begin{tabular}{c S[table-format=5.2(4)] S[table-format=4.0(2)]}
      \hline
        \textbf{Spool} & \textbf{$A$ ($\times 10^{8}$\si{\per\second})} & \textbf{$\lambda_C$ (\si{\nano\meter})}\\
        \hline
        {G652B 20km} & 19900(1460) &  1620(5) \\
        {G652D 20km} &  3450(341) & 1730(8) \\
        {G652B 3km} & 20.3(8.35)& 2070(45) \\ 
        {G652D 2km} & 123(93) & 1860(65)\\
        \hline
        {Theory} 540km & 208400(1400) & 1549(1)  \\
        {Theory} 180km & 120300(610) & 1567(1)  \\
        {Theory} 60km & 56400(230) & 1581(1)  \\
        {Theory} 20km & 22440(80) &  1589(1)  \\
        {Theory} 2km & 2450(10) &  1592(1) \\
        \hline
        \end{tabular}  
\end{table}

The spectrum of the light emitted (\autoref{Fig:SpecResult}), obtained using a CWDM multiplexer, shows that the emission from the water peaks are a large part of the emitted blackbody spectrum. Importantly, this emitted light is in the 1430-1470\si{\nano\meter} CWDM channels, which are not considered water peak channels in the attenuation spectrum. We assume that the asymmetry in the peak is likely due to preferential reabsorption of the emitted light in the 1390 and 1410\si{\nano\meter} channels (akin to a Stokes shift). Ignoring the additional complexities from the -OH emission, the behaviour roughly follows that of the theoretical spectrum detailed in Figure \ref{Fig:Theory}(h). This reinforces that the spectrum we are measuring is an interplay between the rapidly decreasing detector efficiency at longer wavelengths, and the rapidly increasing blackbody spectrum at longer wavelengths. This roll off behaviour is challenging to characterise using manufacturer specifications, as these devices are not designed to be used in the way in which we are using them, and broadband calibration of this specific detector's quantum efficiency outside the manufacturers range was beyond our resources.

The final measurements of each fibre were taken at temperatures well beyond standard operating temperatures, and as a result 3 out of the 4 optical fibres were damaged due to the melting of the plastic spool. The fibre breakage was characterised by an increase in attenuation from the expected -0.2dB/km. After determining the 20\si{\kilo\meter} G652D fibre was broken, we modified our procedure by first measuring at temperatures lower than 80\si{\celsius}, and by measuring that the attenuation of the fibre was within expected ranges before proceeding to further experiments. As we were not measuring the attenuation of the 20\si{\kilo\meter} G652D, we are treating all those data points as though they were from a broken fibre, though we still present curve fit data for the entire set. We have included data points obtained from possibly broken fibre in \autoref{Fig:BBResult} as unfilled markers because they present an interesting comparison. Note that the broken points of the 2\si{\kilo\meter} G652D and 3\si{\kilo\meter} G652B spools at 120\si{\celsius} and 15\si{\celsius} move closer to the result of the 20\si{\kilo\meter} spools, and further away from the centre of the fit. This similarity suggests that the 20km spool is behaving similarly to optically thick silica glass as predicted and expected, rather than the low temperature (intact) behaviour where the shorter spools, which are effectively transparent in the overlap of the blackbody emission and detection bandwidth (at $\approx$1600\si{\nano\meter}). The implication we take from this is that the optically thick regime $\tau\gg1$ and our experiment are in good agreement, despite our expected optical thickness at 20km being close to 1 in the telecom band.

Using the measurements at 150\si{\celsius} for the 20\si{\kilo\meter} G652B fibre (Figure \ref{Fig:BBResult}), we expect approximately $1300$ photons per second across our entire detector bandwidth. If we sum the measured photon rate of our measured channels in Figure \ref{Fig:SpecResult}, and account for the specified 2.2\si{\decibel} insertion loss of our CWDM mux, we obtain approximately 1200 photons per second. This suggests that approximately 100 counts per second are out of band of our CWDM multiplexer, and are most likely coming from the longer wavelengths. We tabulate this result, along with the theoretical spectrum result, in \autoref{App:Theory}. We expect that the difference between the theoretical and experimental spectrum is compounded by the fact that the water emission peak is proximate to our detectors peak quantum efficiency, whereas the theoretical spectrum is dominated by the combination of the rapidly decreasing quantum efficiency at longer wavelengths, against the increasing emission with higher temperatures.

The theoretical calculations suggest that the spectrum is reasonably consistent across our measured temperature range, thus we can estimate the implications for long distance quantum communication. As a rough estimate, 40\si{\celsius} temperatures may be reached for overhead deployment of optical fibres during summer in warm climates (consider that asphalt can reach 50 to 60\si{\celsius} in summer in hot climates \cite{RN688}). 

Taking the ratio of the rate in the 1550\si{\nano\meter} CWDM channel (Figure \ref{Fig:SpecResult}) and the total measured rate (Figure \ref{Fig:BBResult}) ($\approx10$\%), then multiplying by the total at 40\si{\celsius}($\approx$1.5 \si{\per\second}), we obtain a blackbody rate of approximately 0.15\si{\per\second} in the 1550\si{\nano\meter} CWDM channel (or around 0.1\si{\per\second\per\tera\hertz}), \and approximately 0.01\si{\per\second} at 30\si{\celsius}. We show various background rates in Figure \ref{Fig:RateLoss} to demonstrate the impact to quantum communication based on the work of \cite{RN636}.

The presence of blackbody radiation also has implications for the limits of classical communication through the Shannon-Hartley theorem \cite{RN797}:
\begin{align}
C = B \log _2 \left( 1+ \frac{S}{N}\right)
\end{align}
Where $C$ is the capacity, $B$ is the bandwidth, and $S$ and $N$ are the signal and noise levels respectively. For the 1550\si{\nano\meter} CWDM channel, the bandwidth is approximately 2.2\si{\tera\hertz} (equivalent to 18\si{\nano\meter}). The noise levels are taken to be 0.1 photon/s and 0.01 photon/s for 40\si{\celsius} and 30\si{\celsius} respectively, and the results for the channel capacity is shown in \autoref{Tab:ClassicalLimit}, with signal power levels of 10\si{\milli\watt} and 1\si{\watt}. For comparison to literature, these have been converted to a spectral efficiency in bits/s/Hz. Note that for single mode fibre, the state-of-the-art around 13 bits/s/Hz \cite{RN795, RN794}, and the best case limit in our work is approximately 69 bits/s/Hz. For multimode examples, records are significantly higher at 332 bits/s/Hz \cite{RN796}, however we have not calculated a theoretical background rate for multimode fibres. This suggests that blackbody noise may impose a limit for spectral efficiencies a factor of 5 above the current state of the art, although more accurate measurements would be required to determine this limit, as our 30\si{\celsius} is an extrapolation, slightly beyond our measurement limit. Measurements on multimode fibre would also be insightful.

\begin{table}
\centering
\caption{Shannon-Hartley capacity of a CWDM channel in the presence of blackbody noise, compared to high bandwidth communication records. $\Delta\nu$ is the channel bandwidth, $C$ is the channel capacity or data rate, and $C/\Delta\nu$ is the spectral efficiency.}\label{Tab:ClassicalLimit}
      \begin{tabular}{c c c c}
      \hline
        \textbf{} & \textbf{$\Delta\nu$ [THz]} & \textbf{$C$ [Tbit/s]}& \textbf{$C/\Delta\nu$ [bit/s/Hz]}\\
        \hline
        30°C, 10mW      & 2.2    &   138    & 62.7\\
        40°C, 10mW      &        &    131   & 59.5\\
        30°C, 1W        &        &    152  & 69.1\\ 
        40°C, 1W        &        &    145  & 65.9\\
        \hline
         \cite{RN795} Single-mode & 19.8 & 256.4  & 12.95\\
         \cite{RN794} Single-mode & 16.81 & 178.08  & 10.58 \\
         \cite{RN796} Multi-mode & 4.6 & 1530  & 332 \\
        \hline
        \end{tabular}  
\end{table}

 Using detectors with higher quantum efficiency and lower dark count rates (SNSPDs, etc.), and longer integration time, it’s likely possible to obtain more accurate results with the lock in technique, especially for lower temperatures. This would potentially allow for more detailed measurement of the emission spectrum at lower temperatures than the 40\si{\celsius} we used in our experiment. This would allow confirmation that the extrapolation we are using for 30\si{\celsius} is indeed accurate.

Though our measurements have been made on silica fibres, other fibre technologies will exhibit thermal emission, though the absorption characteristics are greatly different between these technologies. Metal fluoride glasses (such as ZBLAN) are of interest due to lower loss than silica, however we note that it is currently prohibitively expensive for long distance communication, and have absorptive (and thus emissive) loss mechanisms in the deeper infrared \cite{RN781}. Hollow core photonic crystal fibres are also of interest in research \cite{RN782}, though in this case the emissive properties of air will be relevant, and these fibres are also expensive when compared to silica. Silica cored photonic crystal fibres will have similar emission characteristics to step index silica fibres, though particulars of the IR vibrational loss are dependent on manufacturing \cite{RN55}, which will be different for engineering reasons, and the confinement in the core is significantly different to step index fibres, so a different model would be required to predict the blackbody emission behaviour.

\section{Conclusion}
Though the measured amount of blackbody radiation is small in an absolute sense, during long distance quantum communication experiments the signal level is quite small (one example is \cite{RN203} 4.3 photons/s over 192km). This noise is also additive, and occurs at wavelengths which are typically used for telecomm. This means that long distances are not only limited by reduced photon rates through attenuation, but also maximum achievable fidelities of a particular quantum link. There also exists a trade-off, where the ideal attenuation channel (1550nm) has more blackbody noise than channels with greater attenuation (by a factor of 10). There is also a trade off between length and noise, though based on our theoretical model, the 'knee' of the emission against length is likely around 100km (above which noise will only increase by a small amount). In classical communication, blackbody noise is likely to be negligible, except in cases where the Shannon-Hartley capacity is being reached. We expect that this result will likely factor in to design considerations for practical quantum networks in the future, and that further investigation is required in both silica and other fiber technologies to get a more accurate understanding of this noise process, and improved detection is required for obtaining accurate measurements at lower temperatures.

\section*{Acknowledgment}
This research is supported by the Commonwealth of Australia as represented by the Defence Science and Technology Group of the Department of Defence. This research is supported by the Queensland Government via the Queensland Defense Science Alliance. We would like to acknowledge the Queensland Micro- and Nanotechnology Centre for the use of their oven.
\appendix
\section{Theoretical calculations} \label{App:Theory}
\begin{figure*}
    \includegraphics[width=\textwidth]{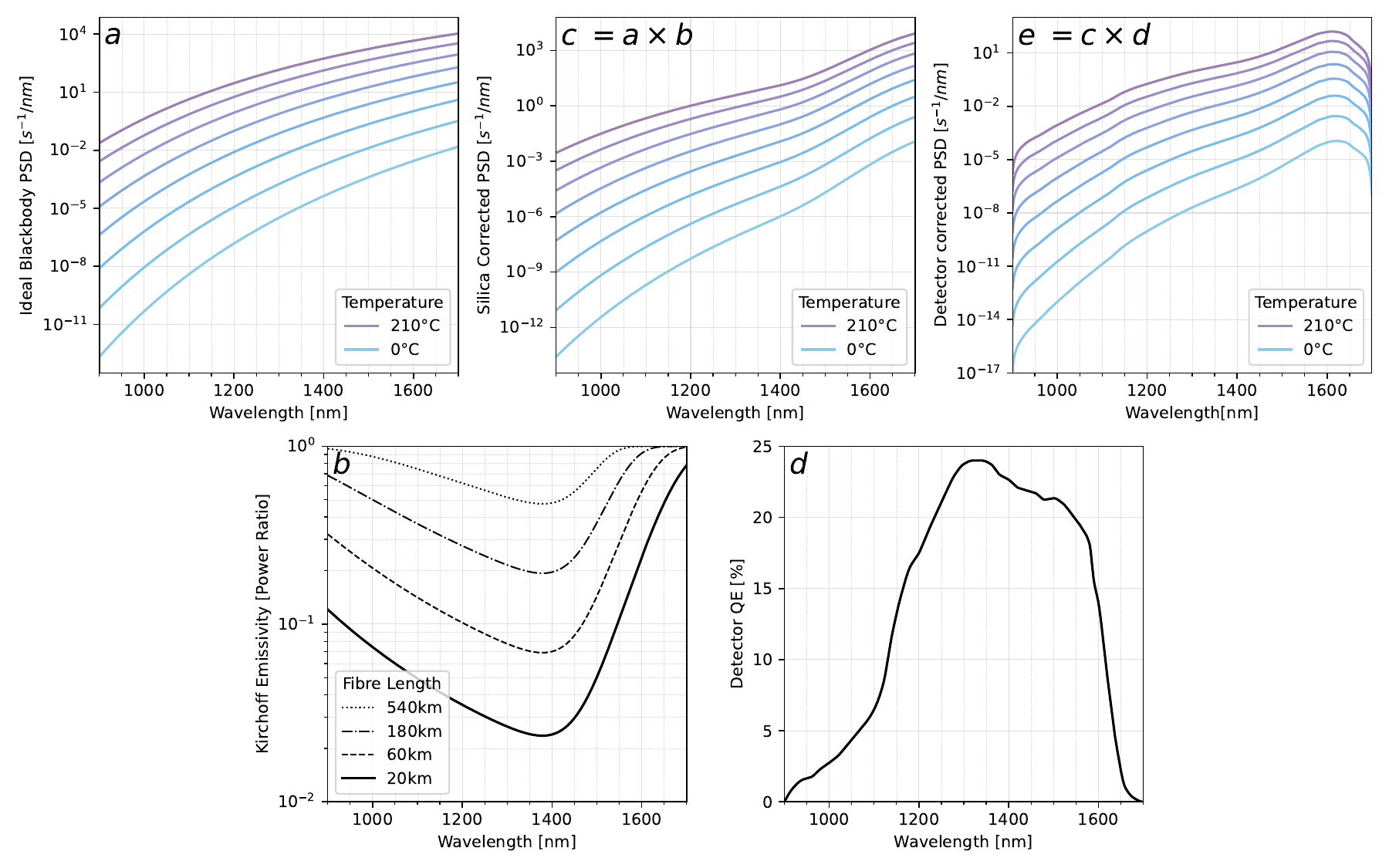}
    \caption{Theoretical calculations of blackbody emitted from an optical fibre. (a) is the ideal blackbody curve of a standard single mode optical fibre, in units of photons per second per nanometer. (b) Is the emissivity calculated using Kirchhoff's law of thermal radiation for several lengths of fibre. (c) is the spectrum (a) accounting for the emissivty for 20km fibre length. (d) is the detector efficiency of the IDQ Qube SPAD. (e) is the spectrum accounting for the detection efficiency, integrating these curves gives an expected signal level for a given temperature.}
    \label{Fig:Theory}
\end{figure*}
To use Planck's law, we need to have both a blackbody surface and an angle of emission. The blackbody surface is taken to be the cross sectional area of the core of a SMF28e fibre, which is 8.2$\mu$m. The solid angle can be calculated using the standard spherical cap calculation $\Omega=2\pi (1- \cos \alpha)$. As a crude approximation, a ray from inside the core must internally reflect, so must have an angle of incidence greater than the critical angle. By constructing our spherical cap such that the radius of the sphere makes and the diameter of the core meet at the critical angle, we get the conic angle of our spherical cap as $\alpha = \frac{\pi}{2}-\theta_c$. We can then use the spec sheet numerical aperture and core index to obtain $\theta_c$. Applying these geometric constraints to Planck's law gives standard curves for blackbody emission in units of power. We can then convert to photons per second as a power unit through Planck's relation $E=hc/\lambda$. This is shown in Figure \ref{Fig:Theory}(a), focusing on the wavelength range that we will be able to detect.

Using Kirchoff's law of thermal radiation, we know that the emissivity and absorptivity are equal. The attenuation spectrum of silica fibre is well discussed in literature \cite{RN55}, and we use the IR vibrational absorption and Urbach absorption.  Note that manufacturing and composition changes can change the absorption characteristics of both Urbach and IR vibration absorption. This attenuation is calculated for 20km, then converted into a power ratio. This gives us Figure \ref{Fig:Theory}(b). By multiplying the emission spectrum by the emissivity, we obtain the emissivity corrected curve show in Figure \ref{Fig:Theory}(c).

The ID Quantique Qube NIR SPAD has an indicative quantum efficiency spectrum given by the manufacturer, which we have traced for the curve in Figure \ref{Fig:Theory}(d). By multiplying the emissivity corrected curve by the efficiency spectrum, we obtain theoretical detection spectrum shown in Figure \ref{Fig:Theory}(e) because of our unit conversion from power to photons/second. 

To obtain the expected blackbody rate for a given temperature, we integrate across the detection range and obtain Figure \ref{Fig:TheoryCurves}. Only the 20km data points are shown, with curve fits shown for 20, 60, 180 and 540km of fibre. These integrals have a built in assumption that the detection efficiency outside the 900-1700nm range is zero, though in reality the behaviour is likely non-trivial. We fit to an exponential function which matches the exponential in Planck's law:
\begin{figure}
    \includegraphics[width=\linewidth]{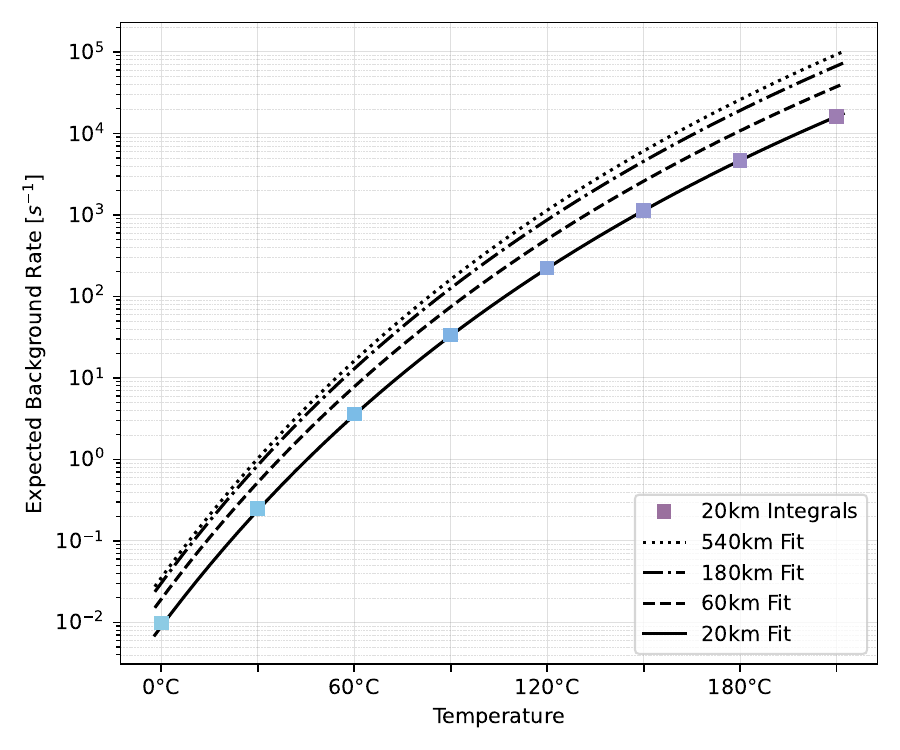}
    \caption{Theoretical signal level for a variety of temperatures and fibre lengths. The square markers indicate the integration results for the curves presented in \autoref{Fig:Theory} (e), for comparison to the measured results. Longer lengths have their markers omitted for clarity.}
    \label{Fig:TheoryCurves}
\end{figure}
\begin{figure}
    \includegraphics[width=\linewidth]{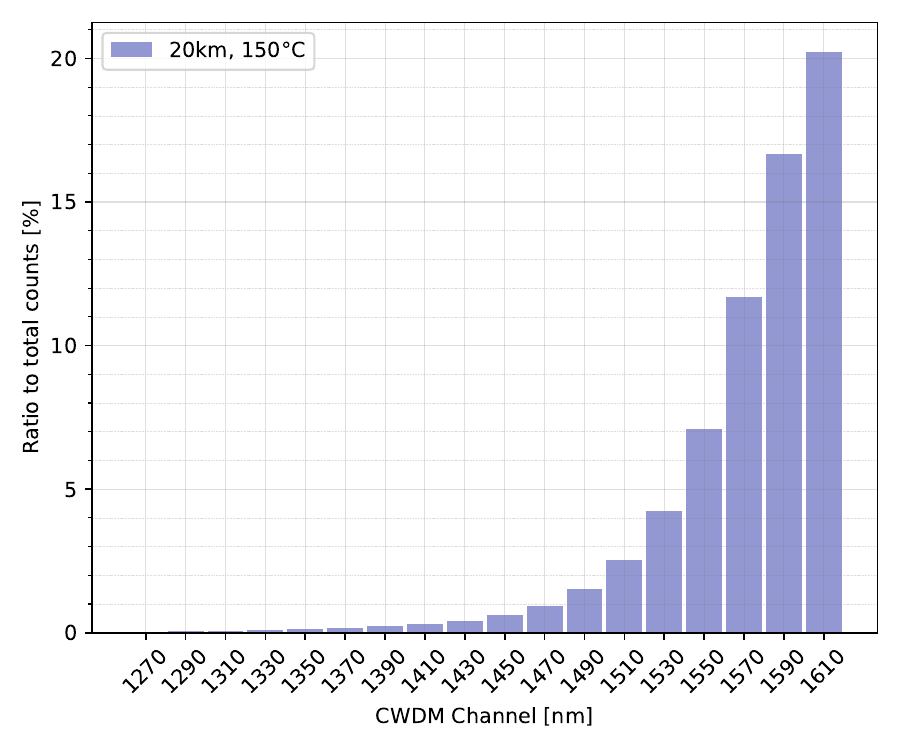}
    \caption{Theoretical proportion of signal in each CWDM channel compared to the total signal at 150°C for 20km of fibre, to compare to \autoref{Fig:SpecResult}.}
    \label{Fig:CWDMSpectrum}
\end{figure}
\begin{align}
    R=A e^{\frac{hc}{k_B}\frac{1}{\lambda_c T}}
\end{align}
The curve fit function gives magnitudes and associated errors as shown in the lower half of \autoref{Tab:CFResults}. We can also use integrals to determine what our expected signal levels would be in CWDM channels to match with Figure 4. These are presented as a percentage of the signal in the 1610nm channel in Figure \ref{Fig:CWDMSpectrum} for 20km of fibre at a temperature of 150°C. We also present this distribution for several lengths and temperatures in \autoref{Tab:CWDMInfo}, as well as the calculated values from the experimental result presented in \autoref{Fig:SpecResult}.

\begin{table}
\centering
\caption{ Theoretical proportion of signal in each CWDM channel based on integrating the results in \autoref{Fig:TheoryCurves} for ideal filtering, compared to the experimental result in \autoref{Fig:SpecResult}. Note that '---' indicates data not measured, and the out of band calculation includes all '---' data.}\label{Tab:CWDMInfo}
      \begin{tabular}{S[table-format=4.0]|S[table-format=2.3] S[table-format=2.3] S[table-format=2.3] S[table-format=2.3] |S[table-format=2.3]}
	\hline
								&\multicolumn{4}{S|}{{\textbf{Theory [\%]}}}				& {\textbf{Exp. [\%]}}\\
	{CWDM} 					& {L=180\si{\kilo\meter}}	& {60\si{\kilo\meter}} 	& {20\si{\kilo\meter}} 		& {20\si{\kilo\meter}}	& {20\si{\kilo\meter}}\\
	$\lambda$ [\si{\nano\meter}] 	& {T=30\si{\celsius}} 		& {30\si{\celsius}} 			& {30\si{\celsius}}		& {150\si{\celsius}}	& {150\si{\celsius}} \\
	\hline
	{Below}					& 0.0120		& 0.0072		& 0.0055	& 0.0738	& {---} \\
	1270					& 0.0094		& 0.0056		& 0.0042	& 0.0375	& 0.92 \\
	1290					& 0.0156		& 0.0093		& 0.0070	& 0.0528	& {---} \\
	1310					& 0.0252		& 0.0149		& 0.0113	& 0.0722	& 0.38 \\
	1330					& 0.0395		& 0.0233		& 0.0176	& 0.0967	& {---} \\
	1350					& 0.0610		& 0.0359		& 0.0272	& 0.1283	& 0.64 \\
	1370					& 0.0924 		& 0.0543		& 0.0411	& 0.1677	& {---} \\
	1390					& 0.1406 		& 0.0827		& 0.0625	& 0.2215	& 1.13 \\
	1410					& 0.2175 		& 0.1283		& 0.0971	& 0.2997	& 1.35 \\
	1430					& 0.3461		& 0.2052		& 0.1556	& 0.4202	& 10.22 \\
	1450					& 0.5722 		& 0.3427		& 0.2608	& 0.6183	& 6.24 \\
	1470					& 0.9691 		& 0.5903		& 0.4519	& 0.9440	& 6.15 \\
	1490					& 1.7054 		& 1.0667		& 0.8242	& 1.5217	& 5.06 \\
	1510					& 3.0436 		& 1.9803		& 1.5520	& 2.5427	& 6.69 \\
	1530					& 5.2859 		& 3.6407		& 2.9143	& 4.2502	& 7.62 \\
	1550					& 8.8522 		& 6.5982		& 5.4466	& 7.0918	& 10.81 \\
	1570					& 14.0647		& 11.6334		& 10.0292	& 11.6950	& 11.26 \\
	1590					& 18.5285 		& 17.3925		& 15.9034	& 16.6838	& 12.11 \\
	1610					& 19.9815 		& 21.6307		& 21.3864	& 20.2392	& 11.56 \\
	{Above}					& 26.0376 		& 34.5577		& 40.8020	& 32.8426	&{---} \\	
	\hline
	{Totals}					&				&				&			&			&	\\
	{In band}					& 73.95			&65.44			& 59.19		& 67.08		& 92.14\\
	{Out of band}				& 26.05			&34.56			& 40.81		& 32.92		& 8.86 \\
        \end{tabular}  
\end{table}

There are several limitations to this model. Firstly, the propagation behaviour is more complex than a single wavelength undergoing total internal reflection. Secondly, the behaviour of the Urbach edge is not well understood for the longer wavelengths where we are using it (ie: greater than 0.5\si{\micro\meter}), and has temperature dependence which we have not accounted for. Finally, we have assumed the behaviour of our detectors outside the specified range is zero, which is likely not true. We believe that this model is a good starting point to understand the noise process, rather than a definitive model of the behaviour.
\end{document}